\documentclass[a4paper,fleqn,usenatbib]{mnras}

\usepackage[T1]{fontenc}
\usepackage{ae,aecompl}

\usepackage{graphicx}	
\usepackage{amsmath}	
\usepackage{amssymb}	

\newcommand{\gp}{{\gamma^\prime}}

\newcommand{\dD}{{\delta_{\rm D}}}
\newcommand{\tp}{t^\prime}

\title[Very hard electron spectrum in fast-cooling regime]{Formation of very hard electron and gamma-ray spectra of flat spectrum radio quasar in fast-cooling regime}
\author[D. H. Yan, L. Zhang, and S. N. Zhang]{Dahai Yan$^{1,2}$\thanks{E--mail: yandahai@ihep.ac.cn}, Li Zhang$^{2}$\thanks{E--mail: lizhang@ynu.edu.cn}, Shuang-Nan Zhang$^1$\thanks{E--mail: zhangsn@ihep.ac.cn}\\
$^1$Key Laboratory of Particle Astrophysics, Institute of High Energy Physics,
Chinese Academy of Sciences, Beijing 100049, China\\
$^2$Department of Astronomy, Key Laboratory of Astroparticle Physics of Yunnan Province, Yunnan University, Kunming, 650091, China}

\date{Accepted XXX. Received YYY; in original form ZZZ}

\pubyear{2015}

\begin{document}
\label{firstpage}
\pagerange{\pageref{firstpage}--\pageref{lastpage}}
\maketitle

\begin{abstract}
In external Compton scenario, we investigate the formation of the very hard electron spectrum in the fast-cooling regime, using a time-dependent emission model. It is shown that a very hard electron distribution $N'_{\rm e}(\gp)\propto\gp^{-p}$ with the spectral index $p\sim1.3$ is formed below the minimum energy of injection electron when inverse Compton scattering takes place in the Klein-Nishina regime, i.e., inverse Compton scattering of relativistic electrons on broad-line region radiation in flat spectrum radio quasars. This produces a very hard gamma-ray spectrum, and can reasonably explain the very hard \emph{Fermi}-LAT spectrum of the flat spectrum radio quasar 3C 279 during the extreme gamma-ray flare in 2013 December.
\end{abstract}


\section{Introduction}
Leptonic models have met with considerable successes in
modelling the broadband (from radio to $\gamma$-ray frequencies) spectral
energy distribution (SED) of all classes of blazars \citep[e.g.,][]{ghisellini10,zhang12,ghisellini14,yan15c}.
In leptonic models, non-thermal emission is produced by
synchrotron emission of relativistic electrons in a comoving magnetic field and inverse Compton (IC) emission of relativistic electrons on low energy photons. For IC scattering, low-energy seed photons may be provided by synchrotron radiation (synchrotron-self Compton, SSC) and various external radiation fields (EC): (1) accretion disk radiation \citep[EC-disk;][]{dermer93,dermer02a},
(2) broad-line region (BLR) radiation \citep[EC-BLR;][]{Sikora1994}, and (3) dust IR radiation \citep[EC-dust;][]{bl}.
Time-dependent leptonic models have been developed to study the observed variability features \citep[e.g.,][]{Masti,Li,kus00,bott02,bott10,chen12,Saito}.

Relativistic electron energy distribution (EED) is crucial for studying the non-thermal radiation from blazars.
A static broken power-law EED \citep[see][for constraining various EEDs with observations]{Yan13,Zhou14} is frequently used to model the SEDs of blazars \citep[e.g.,][]{Tavecchio,finke08,Man,Yan12,Yan14,zhang14,Kang}.
An initial single power-law electron distribution can be deformed to become a broken power-law distribution due to radiative energy losses.
In short, in the slow-cooling regime (i.e.,the minimum energy of injected EED $\gp_{\rm min}$ less than the broken energy $\gp_{\rm b}$ of cooled EED), the electron spectrum below $\gp_{\rm b}$ has a spectral index $p=s$, where $s$ is the spectral index of the injected single power-law EED; above $\gp_{\rm b}$, the spectrum is softened by cooling, and has an index $s+1$. In the fast-cooling regime (i.e., $\gp_{\rm b}=\gp_{\rm min}$), $p\sim2$, independent of $s$; above $\gp_{\rm min}$, the spectrum also has an index $s+1$ \citep[e.g.,][]{dm09,finke13}. Note that standard shock
acceleration theories predict $s\sim2$.
Modelling the SEDs of \emph{Fermi}-LAT detected blazars \citep{LBASsed} returns a standard electron spectrum with $p\sim2$ below $\gp_{\rm b}$ \citep[e.g.,][]{Yan12,Yan14,zhang12,Kang}.

However, the standard picture mentioned above faces challenges when trying to explain the very hard TeV emission
detected for several high-synchrotron-peaked BL Lac objects (HSPs), e.g., 1ES 1101-232 and 1ES 0229+200.
Various approaches have been proposed to explain the very hard TeV spectrum of HSPs: the leptonic models in extreme regime
\citep[e.g,][]{Kata,Tavecchio09}, the modified leptonic models in normal regime \citep[e.g.,][]{Lefa,bott08,Yan12b}, and the Ultra-high energy cosmic rays propagation models \citep[e.g.,][]{essey11,murase12,Yan15b}.

The recently observed very hard GeV spectrum of the flat spectrum radio quasar (FSRQ) 3C 279 during an extreme $\gamma$-ray flare \citep{Hayashida15} also challenges the standard picture. Modeling by \citet{Hayashida15} with a broken power-law EED showed that a very hard electron spectrum
with $p\sim1$ below $\gp_{\rm b}\sim3000$ is required to explain this very hard \emph{Fermi}-LAT spectrum.
In the slow-cooling regime, such a hard emitting electron spectrum requires a very hard injection electron distribution. Several mechanisms have been proposed to produce a very hard injection electron distribution, for example,  magnetic reconnection \citep[e.g.,][]{zh,Guo14,Guo15,Sironi,Werner} and relativistic shock \citep[e.g.,][]{Stecker}.
\citet{Asano} recently explained the very hard GeV spectrum in a stochastic acceleration model.

Here we study the formation of the very hard electron spectrum of FSRQ in the fast-cooling regime.
We show that a very hard electron spectrum with $p\sim1.3$ below minimum injection energy is produced in the fast-cooling regime owing to IC scattering on BLR radiation in the Klein-Nishina (KN) regime. This produces very hard EC components, which can reasonably account for the very hard \emph{Fermi}-LAT spectrum of 3C 279. In Section~\ref{model}, we describe our model; numerical results are showed in Section~\ref{NR}. In Section~\ref{app}, we apply our approach to the very hard \emph{Fermi}-LAT spectrum of 3C 279. In Section~\ref{sd}, we give summary and discussion.
Throughout the paper, we use cosmology parameters $H_0=71\rm \ km\ s^{-1}\ Mpc^{-3}$, $\Omega_{\rm m}=0.27$, $\Omega_{\Lambda}=0.73$.

\section{Model}
\label{model}

\subsection{Model Description}

In a one-zone leptonic model, it is assumed that the emission is produced by relativistic electrons injected in
a homogeneous blob of comoving radius $R'$. The emission blob moves with relativistic speed (corresponding to the bulk Lorentz factor $\Gamma$) towards us. Due to the beaming effect, the observed emission is strongly boosted. For a blazar, we assume the Doppler factor $\delta_{\rm D}=\Gamma$. Note that quantities in the frame comoving with the jet blob are primed.

Relativistic electrons lose energy due to synchrotron and IC radiation.
The kinetic equation governing the temporal evolution of the
electrons distribution $N_e'(\gp,\tp)$ is
\begin{equation}
\frac{\partial N_e'(\gp,\tp)}{\partial\tp}=\frac{\partial}{\partial\gp}[\dot{\gp}N_e'(\gp,\tp)]-\frac{N_e'(\gp,\tp)}{\tp_{\rm esc}}+Q'(\gp,\tp)\ ,
\label{eed}
\end{equation}
where $N'_e$ is the differential electron number and $\tp_{\rm esc}$ the escape timescale.
$\dot{\gp}$ is the total cooling rate, and $Q'(\gp,\tp)$  is the electron injection rate.

We take into account radiative cooling due to synchrotron radiation, SSC and EC. Therefore, we have
\begin{equation}
\dot{\gp} (r) = \dot{\gp}_{\rm syn} + \dot{\gp}_{\rm SSC} + \dot{\gp}_{\rm EC} (r).
\end{equation}

The synchrotron cooling rate is given by
\begin{equation}
\dot{\gp}_{\rm syn}=\frac{4c\sigma_{\rm T}}{3m_{\rm e}c^2}u_{\rm B}\gp^2\ ,
\end{equation}
where $$u_{\rm B}=\frac{B^{\prime 2}}{8\pi}$$
is the magnetic energy density and $B'$ the comoving magnetic field, $c$ is the speed of light, $m_{\rm e}$ is the electron mass, and $\sigma_{\rm T}$ is the Thomson cross section.

The SSC cooling rate using
the full KN cross section is \citep[e.g.,][]{Jones68,bott97,finke08}
\begin{equation}
\dot{\gp}_{\rm SSC} = \frac{3 \sigma_{\rm
T}}{8m_{\rm e}c}\int^{\infty}_{0}d\epsilon^{\prime}
             \frac{u_{\rm syn}^\prime(\epsilon^{\prime} )}{\epsilon^{\prime 2}}\ G(\gamma^{\prime}\epsilon^{\prime})\,
\end{equation}
where
$$G(E) =
\frac{8}{3}E\frac{1+5E}{(1+4E)^2}-\frac{4E}{1+4E}\left(\frac{2}{3}+\frac{1}{2E}+\frac{1}{8E^2}\right)$$
$$+\ln(1+4E)\left(
1+\frac{3}{E}+\frac{3}{4}\frac{1}{E^2}+\frac{\ln[1+4E]}{2E}-\frac{\ln[4E]}{E}
\right)$$ $$-\frac{5}{2}\frac{1}{E}\ +\
\frac{1}{E}\sum^{\infty}_{n=1}
\frac{(1+4E)^{-n}}{n^2}-\frac{\pi^2}{6E}-2,$$
and $u_{\rm syn}^\prime (\epsilon^{\prime}) $ is the spectral energy density of synchrotron radiation.

A fairly accurate
approximation for the EC cooling rate, valid in the Thomson through Klein-Nishina regimes, is given by \citet{Moderski}, and it is
\begin{equation}
\dot{\gp}_{\rm EC}=\frac{4c\sigma_{\rm T}}{3m_{\rm e}c^2}u'_{0}\gp^2f_{\rm KN}(4\gp\epsilon'_0)\ ,
\end{equation}
where $u'_{0}$ and $\epsilon'_0$ are the energy density and dimensionless
photon energy, respectively, of the external radiation field in the comoving frame of blob.
The correction function for KN effect is given by
\begin{equation}
f_{\rm KN}(x)=\frac{1}{(1+x)^{1.5}}\ .
\end{equation}

The external radiation includes emissions from broad-line region (BLR) and infrared dust torus.
Their energy densities in the comoving frame as the functions of the distance $r$ from the black hole are given by \citep{Sikora09,Hayashida}
\begin{equation}
u'_{\rm BLR} (r)=\frac{\Gamma^2\tau_{\rm BLR} L_{\rm disk}}{3\pi r^2_{\rm BLR}c[1+(r/r_{\rm BLR})^3]},\
\label{u1}
\end{equation}
\begin{equation}
u'_{\rm dust} (r)=\frac{\Gamma^2\tau_{\rm dust} L_{\rm disk}}{3\pi r^2_{\rm dust}c[1+(r/r_{\rm dust})^4]},\
\label{u2}
\end{equation}
where $\tau_{\rm BLR}$ and $\tau_{\rm dust}$ are the fractions of the disk luminosity reprocessed into BLR radiation
and into dust radiation, respectively. The typical values of $\tau_{\rm BLR}\sim0.1$ \citep[e.g.,][]{ghisellini14} and $\tau_{\rm dust}\sim0.3$ \citep[e.g.,][]{Hao,Malmrose} are adopted in the following calculations.
The sizes of BLR and dust torus are related to the disk luminosity $L_{\rm disk}$ \citep{ghisellini09,ghisellini14}, i.e.,
\begin{equation}
r_{\rm BLR}=10^{17}(L_{\rm disk}/10^{45}\rm \ erg\ s^{-1})^{1/2}\ {\rm cm},
\end{equation}
\begin{equation}
r_{\rm dust}=10^{18}(L_{\rm disk}/10^{45}\rm \ erg\ s^{-1})^{1/2}\ {\rm cm}.
\end{equation}
 Then, we have
\begin{equation}
u'_{\rm BLR} (r)\simeq\frac{0.3\Gamma^2\tau_{\rm BLR}}{1+(r/r_{\rm BLR})^3}\rm \ erg\ cm^{-3},\
\label{u3}
\end{equation}
\begin{equation}
u'_{\rm dust} (r)\simeq\frac{0.003\Gamma^2\tau_{\rm dust}}{1+(r/r_{\rm dust})^4}\rm \ erg\ cm^{-3}. \
\label{u4}
\end{equation}
Therefore, $u'_0$ is also a function of $r$, i.e.,
\begin{equation}
u'_0 (r) = u'_{\rm dust} (r)+u'_{\rm BLR} (r).
\end{equation}

BLR and IR dust radiation is assumed to be a diluted blackbody radiation.
Given that BLR radiation is dominated by $\rm Ly\alpha$ line photons, we adopt an effective temperature for the BLR radiation of $T_{\rm BLR}=4.2\times10^4\ $K, so that the energy density of BLR radiation peaks at
$\approx 2.82 k_{\rm B} T_{\rm BLR}/h \cong 2.5\times 10^{15}$ Hz (corresponding to the mean dimensionless energy $\epsilon_{\rm BLR}=2\times10^{-5}$). We assume an effective temperature for the IR dust radiation of $T_{\rm dust}=1000\ $K \citep{Malmrose}, i.e., the mean dimensionless energy $\epsilon_{\rm dust}=5\times10^{-7}$. Then we have $\epsilon'_0=\Gamma\epsilon_{\rm BLR}$ for $r\leq 2r_{\rm BLR}$, and $\epsilon'_0=\Gamma\epsilon_{\rm dust}$ for $r> 2r_{\rm BLR}$.

We neglect the electron energy loss due to adiabatic expansion,
because in FSRQs the adiabatic cooling with an expanding velocity $\sim0.1c$ is relevant only for very low energy electrons which do not contribute to the radiative output.

\begin{figure} 
\begin{center}
 \includegraphics[width=270pt,height=215pt]{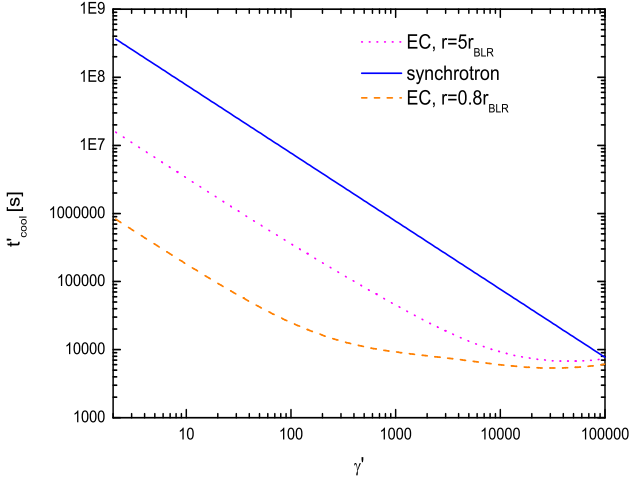}
 \caption{Cooling time for synchrotron radiation and EC processes. $B'=1\ $G and $\Gamma=30$ are used.}
\label{cooling}
\end{center}
\end{figure}

In Fig.~\ref{cooling}, we show the cooling time ($\gp/\dot{\gp}$) for synchrotron radiation and EC processes.
We use $B'=1\ $G, $\Gamma=30$, $\tau_{\rm BLR}=0.1$, $\tau_{\rm dust}=0.3$, and $L_{\rm disk}=1.5\times10^{45}\rm \ erg\ s^{-1}$.
One can find that taking $r=0.8\ r_{\rm BLR}$, significant KN correction takes place at $\gp\gtrsim1/4\Gamma\epsilon_{\rm BLR}\thicksim400$;
taking $r=5\ r_{\rm BLR}$, the KN correction takes place at $\gp>10^4$.

We consider a constant injection during the injection time $\tp_{\rm inj}$. The injection electron distribution is
\begin{equation}
Q'(\gp) = Q'_0\gp^{-s}\rm exp(-\gp/\gp_{\rm cut})\emph{H}\ (\gp;\gp_{\rm min}),
\end{equation}
where $s$ is the spectral index, $\gp_{\rm min}$ is the minimum injection energy, $\gp_{\rm cut}$ is the cut-off energy, and $Q'_0$ [$\rm s^{-1}$] is the normalization constant; $\emph{H}\ (\gp;\gp_{\rm min})=1$ for $\gp>\gp_{\rm min}$, otherwise $\emph{H}\ (\gp;\gp_{\rm min})=0$.

We naturally relate $r$ to time by:
\begin{equation}
r=r_0+c\tp\Gamma,
\label{dis}
\end{equation}
where $r_0$ is the distance where the injection starts.

\section{Numerical Results}
\label{NR}

\begin{figure} 
\begin{center}
 \includegraphics[width=270pt,height=215pt]{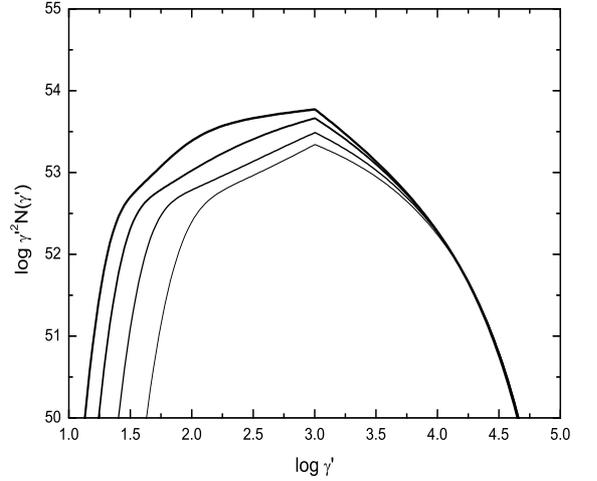}
 \includegraphics[width=270pt,height=215pt]{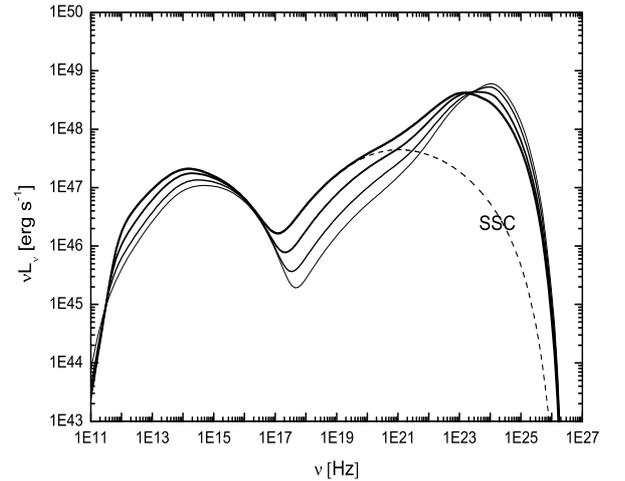}
 \caption{Temporal evolution of EED (top panel) and SED (bottom panel) with $r_0=0.8 r_{\rm BLR}$. The lines from thin to heavy correspond to $\tp=[0.5, 1, 2, 4]\times10^{5}\ $s, respectively. The SSC component is at $\tp=4\times10^{5}\ $s.
 We use $\gp_{\rm min}=10^3$, $\gp_{\rm cut}=10^4$, $Q_0=4.8\times10^{49}\ \rm s^{-1}$, $s=2.1$, $B'=1\ $G, $\delta_{\rm D}=30$, $R'=10^{16}\ $cm, $\tau_{\rm BLR}=0.1$, $\tau_{\rm dust}=0.3$, and $L_{\rm disk}=1.5\times10^{45}\ \rm erg\ s^{-1}$. }
\label{blr}
\end{center}
\end{figure}

\begin{figure} 
\begin{center}
 \includegraphics[width=270pt,height=215pt]{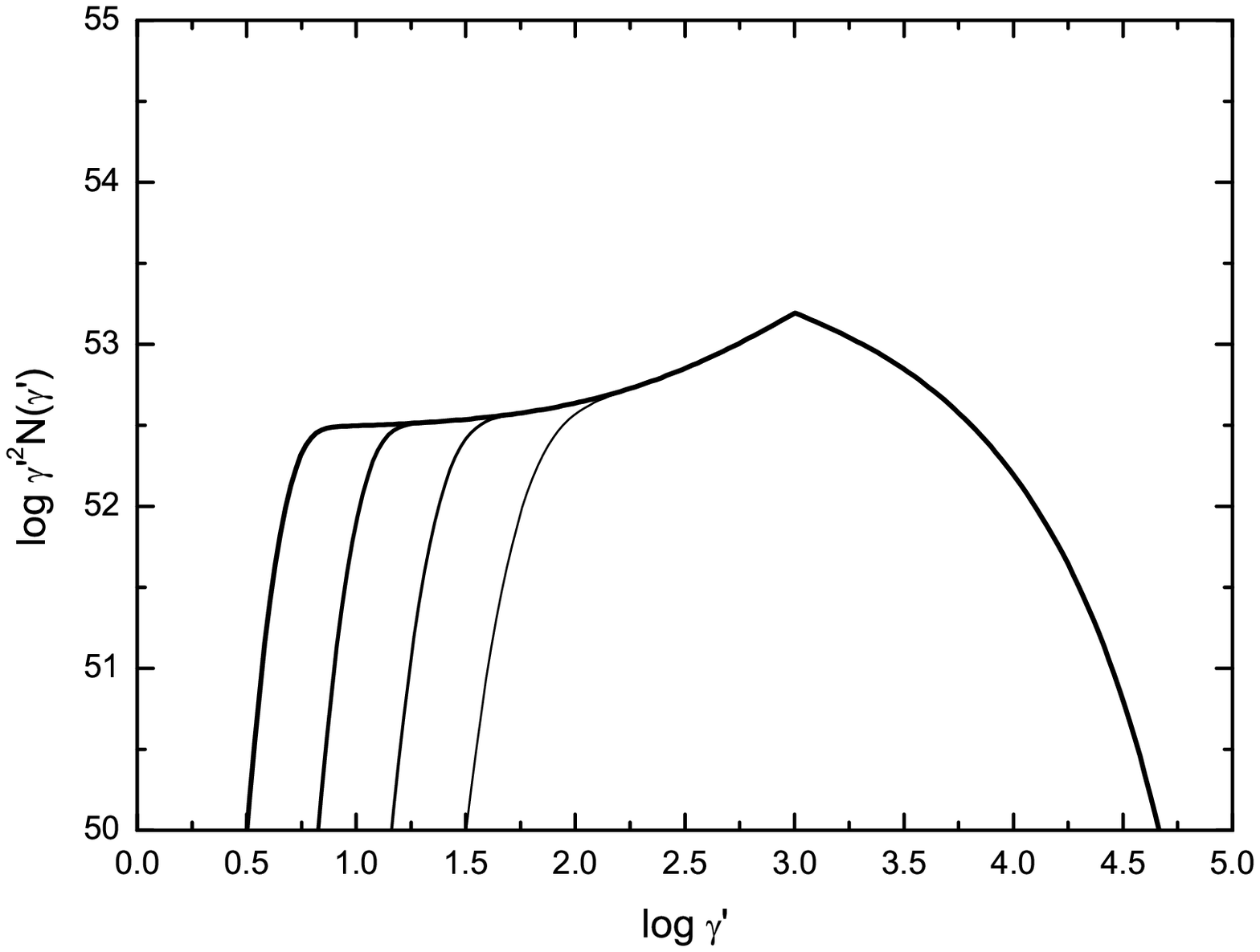}
 \includegraphics[width=270pt,height=215pt]{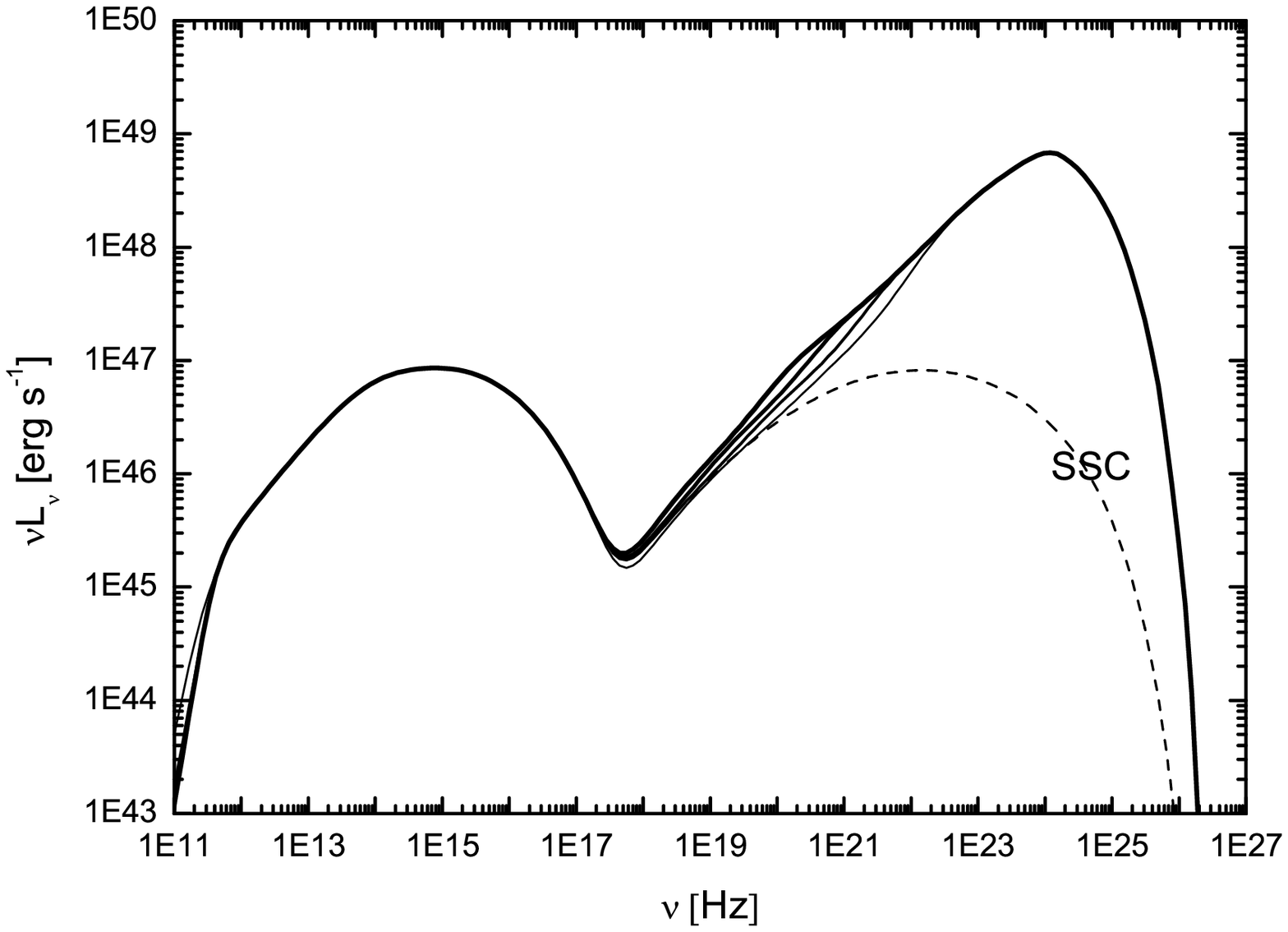}
 \caption{Same as Fig.~\ref{blr}, but with $r=r_0=0.8 r_{\rm BLR}$.}
\label{r0}
\end{center}
\end{figure}

\begin{figure} 
\begin{center}
 \includegraphics[width=270pt,height=215pt]{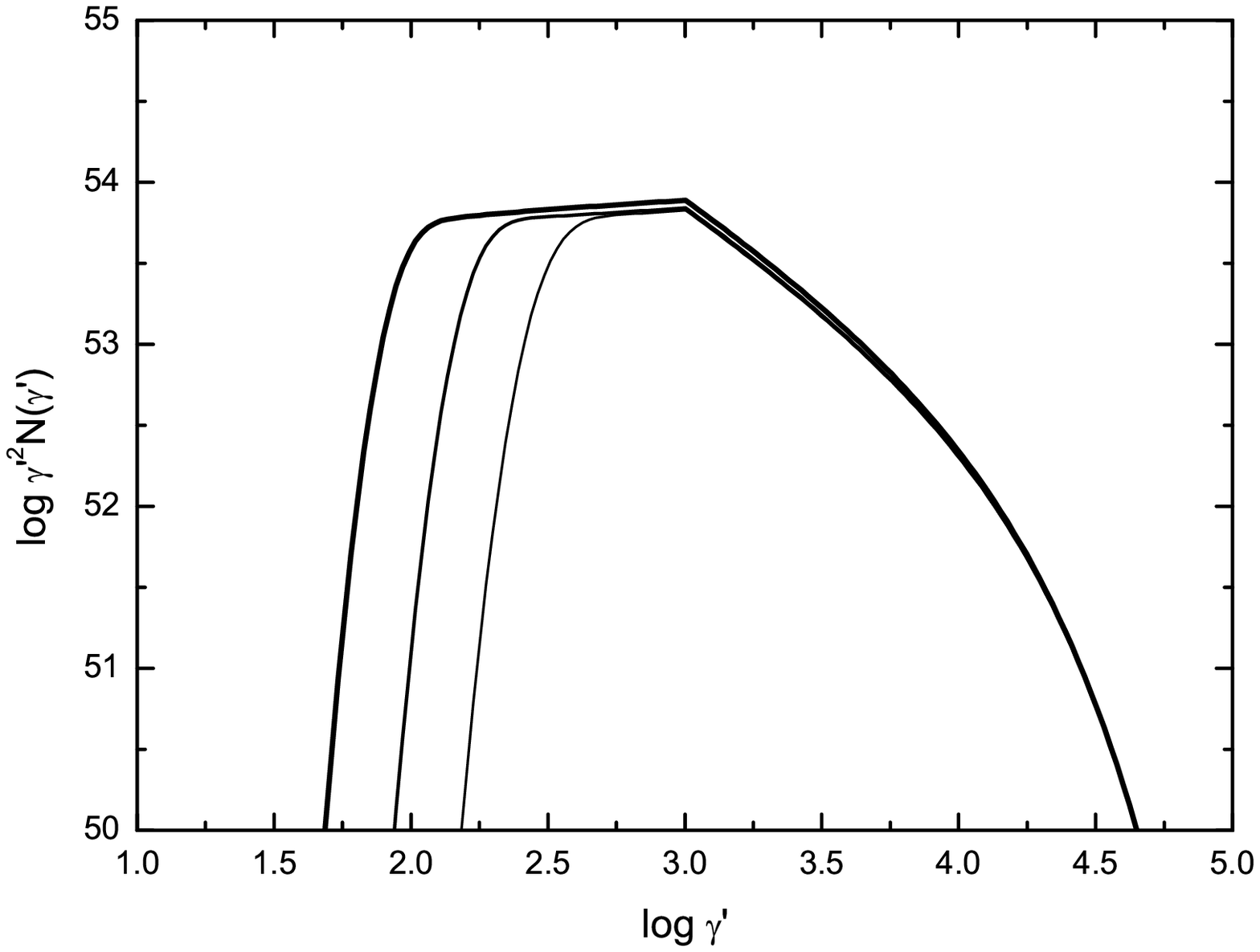}
 \includegraphics[width=270pt,height=215pt]{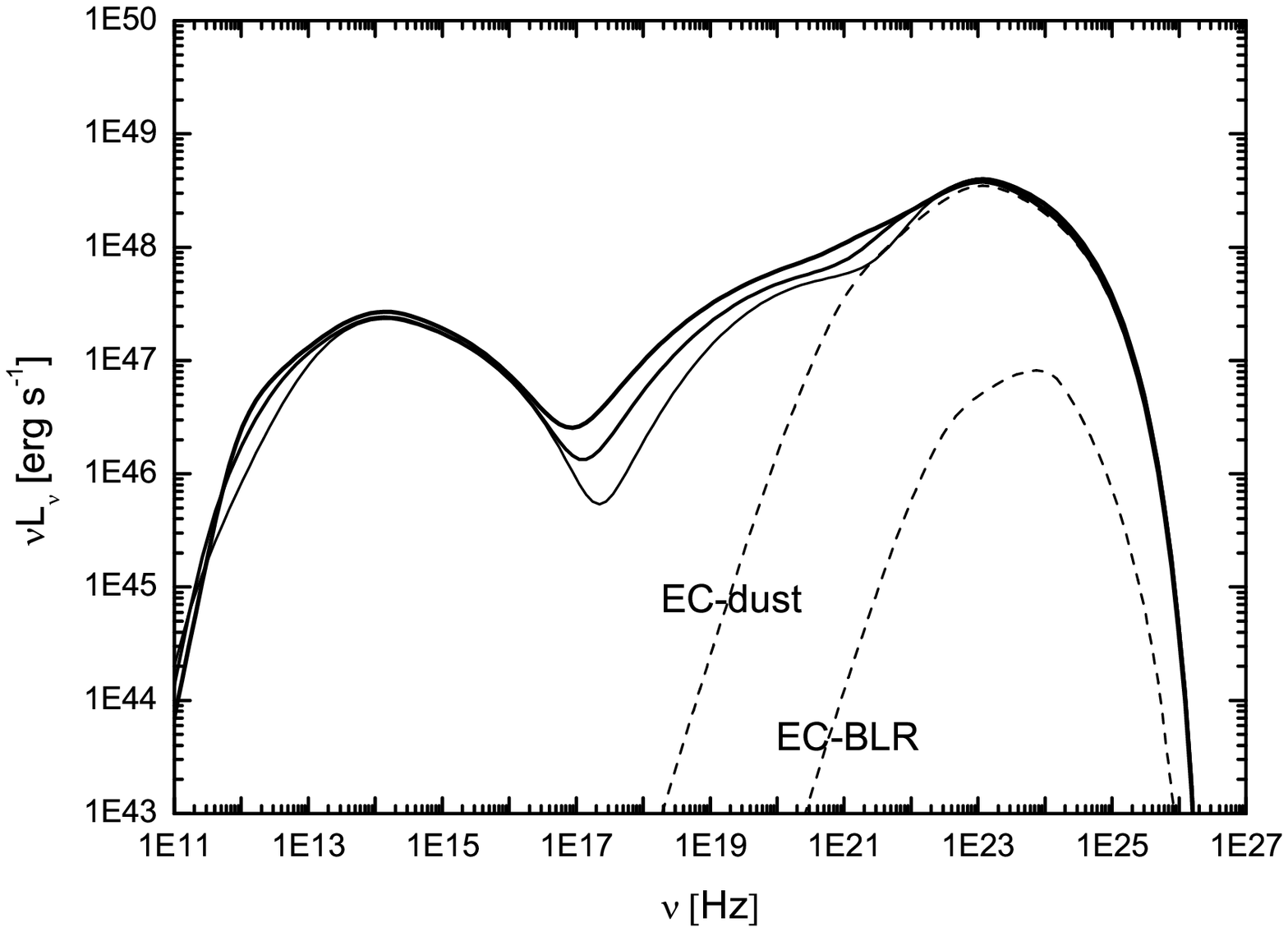}
 \caption{Temporal evolution of EED (top panel) and SED (bottom panel) with $r_0=5 r_{\rm BLR}$. The lines from thin to heavy correspond to $\tp=[1, 2, 4]\times10^{5}\ $s, respectively. The other parameters are same as those in Fig.~\ref{blr}.}
\label{ir}
\end{center}
\end{figure}
We numerically solve equation~(\ref{eed}), adopting the numerical method given by \citet{Chiaberge}.
In the calculations, we use $\tp_{\rm esc}=\tp_{\rm inj}=10^8\ $s.
We calculate the synchrotron and IC spectra using the methods given by \citet{dm09}.
Synchrotron-self absorption is taken into account.

\subsection{Results in fast-cooling regime}

In Fig.~\ref{blr}, we show the temporal evolution of EED and SED in the case of the emission region initially located inside BLR.
One can see that EED develops a very hard $N'_{\rm e}(\gp)\propto\gp^{-1.3}$ form below the minimum injection energy.
This very hard spectrum is different from the standard shape of $N'(\gp)\propto\gp^{-2}$ expected in the case of Thomson or synchrotron cooling processes of the form $\dot{\gp}\propto\gp^2$. The hardening in the electron spectrum is mainly owing to KN energy losses on the BLR radiation. The minimum energy $\gp_1$ of the emitting electron distribution is affected by the evolution time $\tp$ when $\tp<\tp_{\rm esc}$, which can be evaluated by the relation $\tp_{\rm cool}=\tp$.
In the fast-cooling regime, we have $\gp_1<\gp_{\rm min}$.
The $\gamma$-ray spectrum in Fig.~\ref{blr} is the sum of EC-BLR and EC-dust components.
Below $\sim10^{24}\ $Hz, we have a very hard $\gamma$-ray spectrum. When the emission region moves outside BLR, i.e., $\tp>10^{5}\ $s, a softening in EED and SED occurs.

One can see that the variation in EED with time is significant below $\gp_{\rm cut}$ (see top panel in Fig.~\ref{blr}).
The change in EED leads to the change in synchrotron spectrum below $\sim4\times10^6B'\gp^2_{\rm cut}\dD\ $Hz,
and the change in EC spectrum below $\sim10^{20}\epsilon'_0\dD\gp^2_{\rm cut}\ $Hz.
Note that EC-BLR spectrum above $\sim10^{20}\epsilon'_0\dD\gp^2_{\rm KN}\sim2.4\times10^{23}\ $Hz is suppressed by the KN effect, where $\gp_{\rm KN}\sim1/4\epsilon'_0$.
The variation of the X-ray spectrum is due to the change of $\gp_1$.

In above calculations, the energy density of BLR varies with time because $r\sim r_{\rm BLR}$.
In Fig.~\ref{r0}, we show the results for a constant energy density of BLR (corresponding to $r\ll r_{\rm BLR}$) by fixing $r=r_0$.
In this case the electron distribution above $\gp_{\rm min}$ does not vary with time (see top panel in Fig.~\ref{r0}).
Below $\gp_{\rm min}$, the distribution obviously hardens at $\sim1/4\Gamma\epsilon_{\rm BLR}$.
When $\gp_1\ll1/4\Gamma\epsilon_{\rm BLR}$, there is a clear shape $N'_{\rm e}(\gp)\propto\gp^{-2}$ between $\gp_1$ and $\gp\sim100$.
Looking at the evolution of SED,
the multifrequency emissions are weakly variable.

In Fig.~\ref{ir}, we show the temporal evolution of EED and SED in the case of the emission region located far beyond the BLR.
As expected, a standard shape $N_{\rm e}'(\gp)\propto\gp^{-2}$ is formed below $\gp_{\rm min}$.
The $\gamma$-ray spectrum is also softer than that in Fig.~\ref{blr}.

\subsection{Results in slow-cooling regime}

\begin{figure} 
\begin{center}
 \includegraphics[width=270pt,height=215pt]{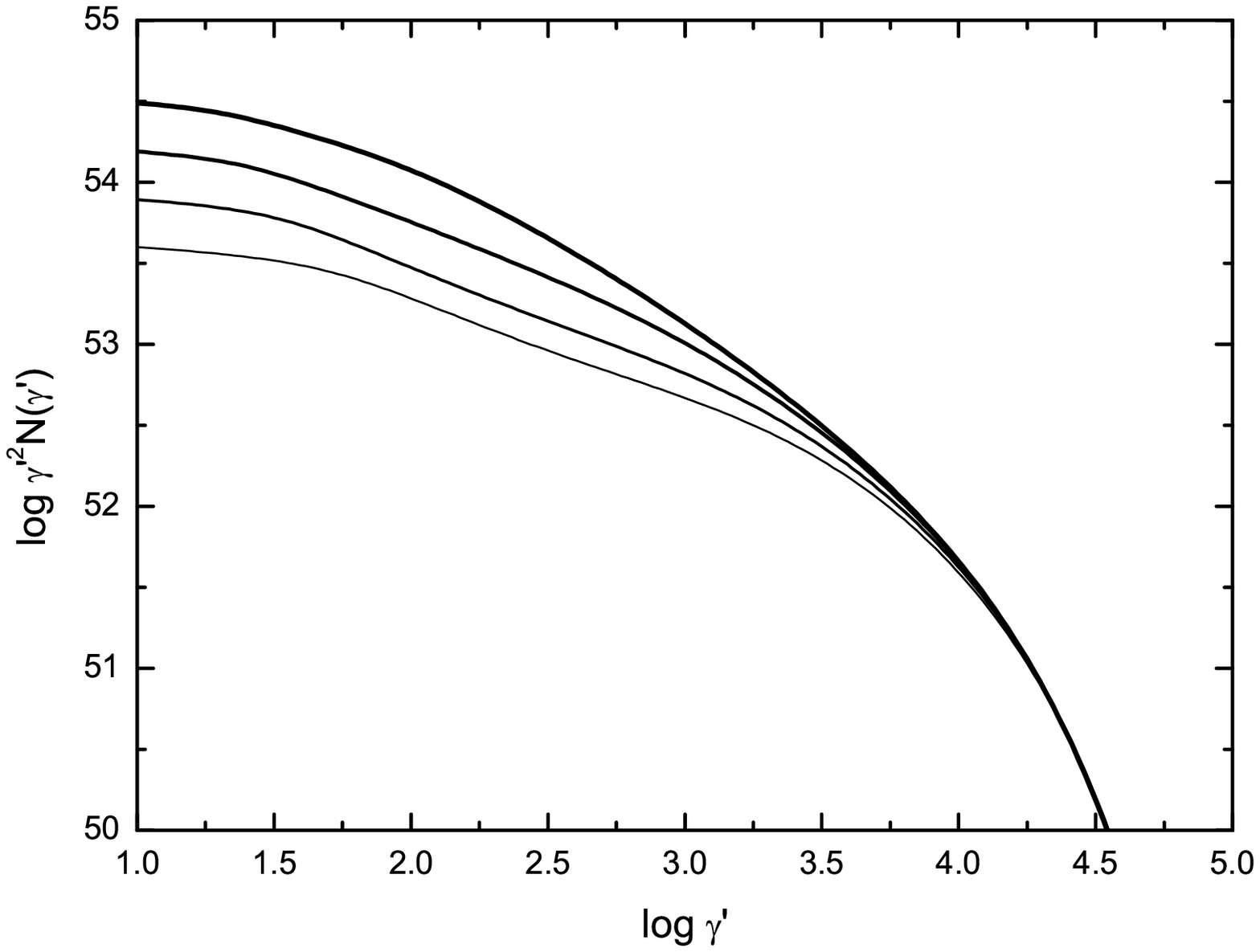}
 \includegraphics[width=270pt,height=215pt]{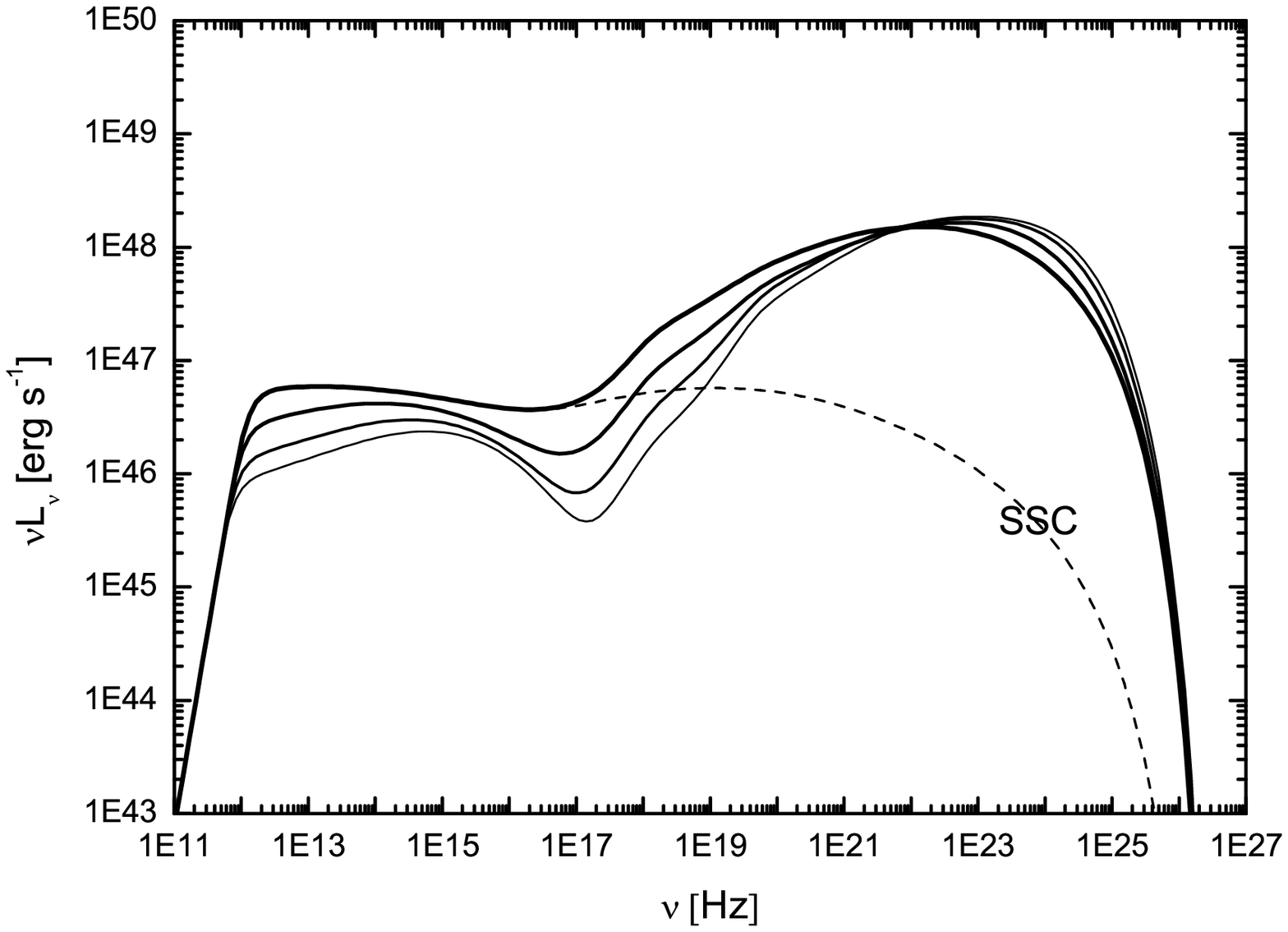}
 \caption{Same as Fig.~\ref{blr}, but with $\gp_{\rm min}=3$ (corresponding to the slow-cooling scenario) and $Q_0=10^{49}\ \rm s^{-1}$.}
\label{slow}
\end{center}
\end{figure}

\begin{figure} 
\begin{center}
 \includegraphics[width=270pt,height=215pt]{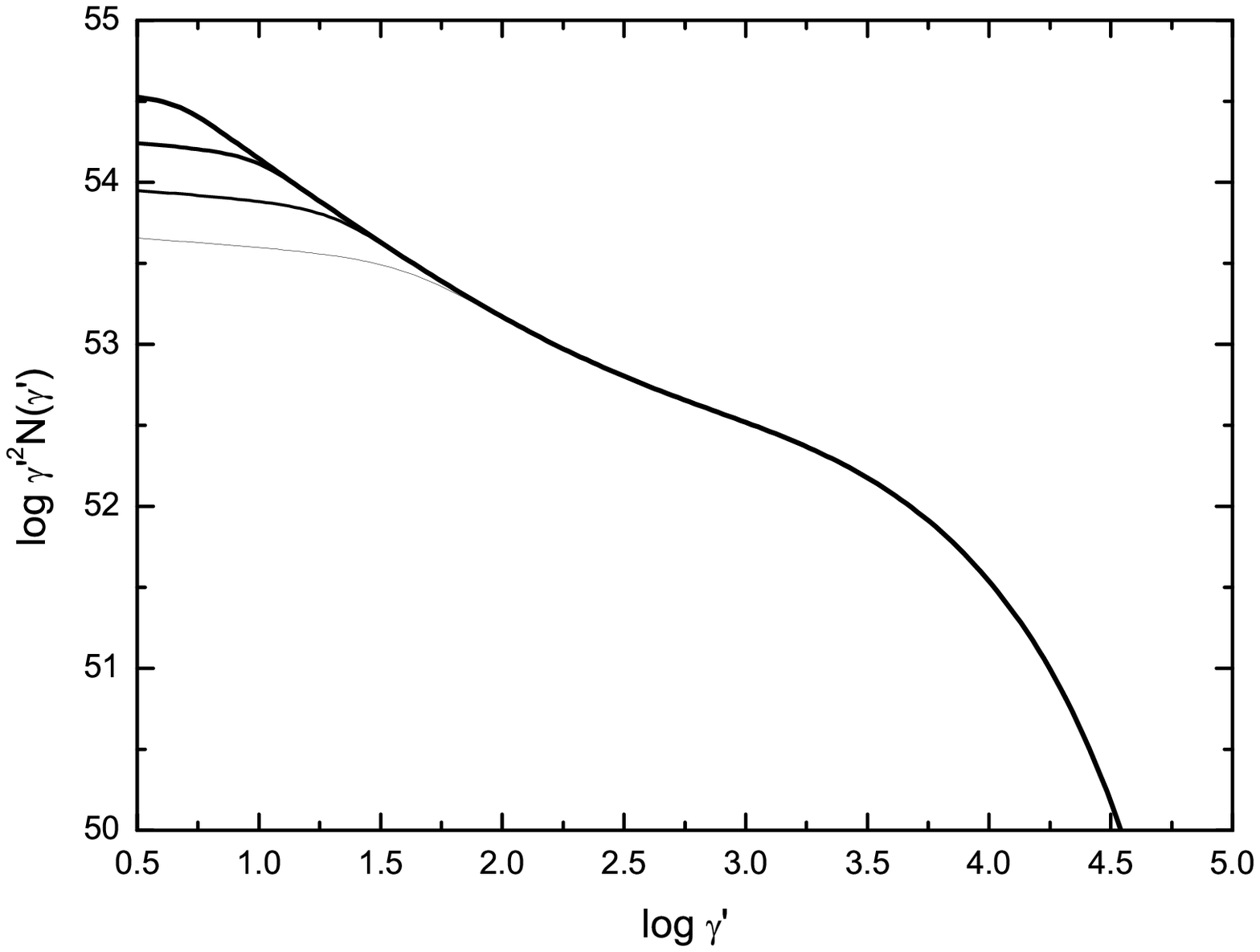}
 \includegraphics[width=270pt,height=215pt]{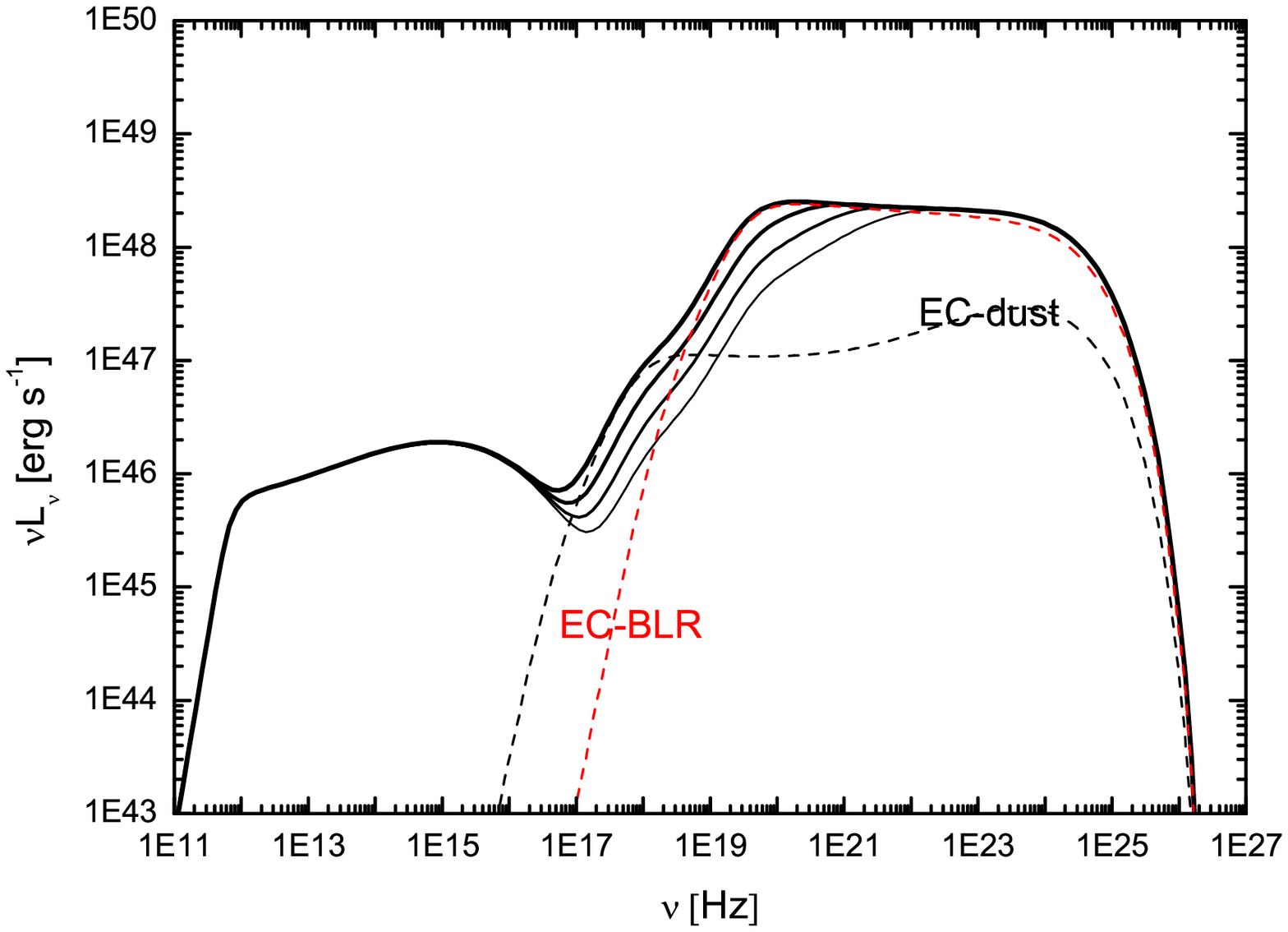}
 \caption{Same as Fig.~\ref{slow}, but with $r=r_0=0.8 r_{\rm BLR}$.}
\label{slow0}
\end{center}
\end{figure}

For comparison, we revisit the KN effect in the slow-cooling region \citep[e.g.,][]{dermer02,kt05,Geo06}.
In Figs.~\ref{slow} and \ref{slow0}, we show the results in the slow-cooling regime;
no very hard electron spectrum with  $p<2$ is seen.
In a constant BLR radiation case (Fig.~\ref{slow0}), it is clear that the EED hardens from $N'_{\rm e}(\gp)\propto\gp^{-(s+1)}$ ($s=2.1$) in the Thomson regime to $N'_{\rm e}(\gp)\propto\gp^{-2.5}$ in the KN regime. This produces a flattening in the EC spectrum (see bottom panel in Fig.~\ref{slow0}). This situation is very similar to the flattening in EC spectrum in FSRQs shown by \citet{Geo06} and the flattening in synchrotron X-ray spectrum in extended \emph{Chandra} jet presented by \citet{dermer02}.

\section{Application to the very hard $\gamma$-ray spectrum of 3C 279}
\label{app}

We apply our approach to the very hard spectrum with photon spectral index $\Gamma_{\gamma}\simeq1.7$ of 3C 279 during an extreme flare \citep{Hayashida15}. The variability timescale of the $\gamma$-ray flare is estimated to be $t_{\rm var}\simeq$2\ hr \citep{Hayashida15}.
This timescale is used to constrain the radius of the emission blob, $R'\lesssim ct_{\rm var}\dD/(1+z)$.
We use $L_{\rm disk}=1.5\times10^{45}\ \rm erg\ s^{-1}$ \citep{yan15}, then we have $r_{\rm BLR}=1.2\times10^{17}\ $cm.
We also assume an inefficient escape, $\tp_{\rm esc}=10^8\ $s\textbf{\footnote{We tested the EED with different $\tp_{\rm esc}$ and found that the escape term is negligible in our model.}}.

\begin{figure} 
\begin{center}
 \includegraphics[width=270pt,height=215pt]{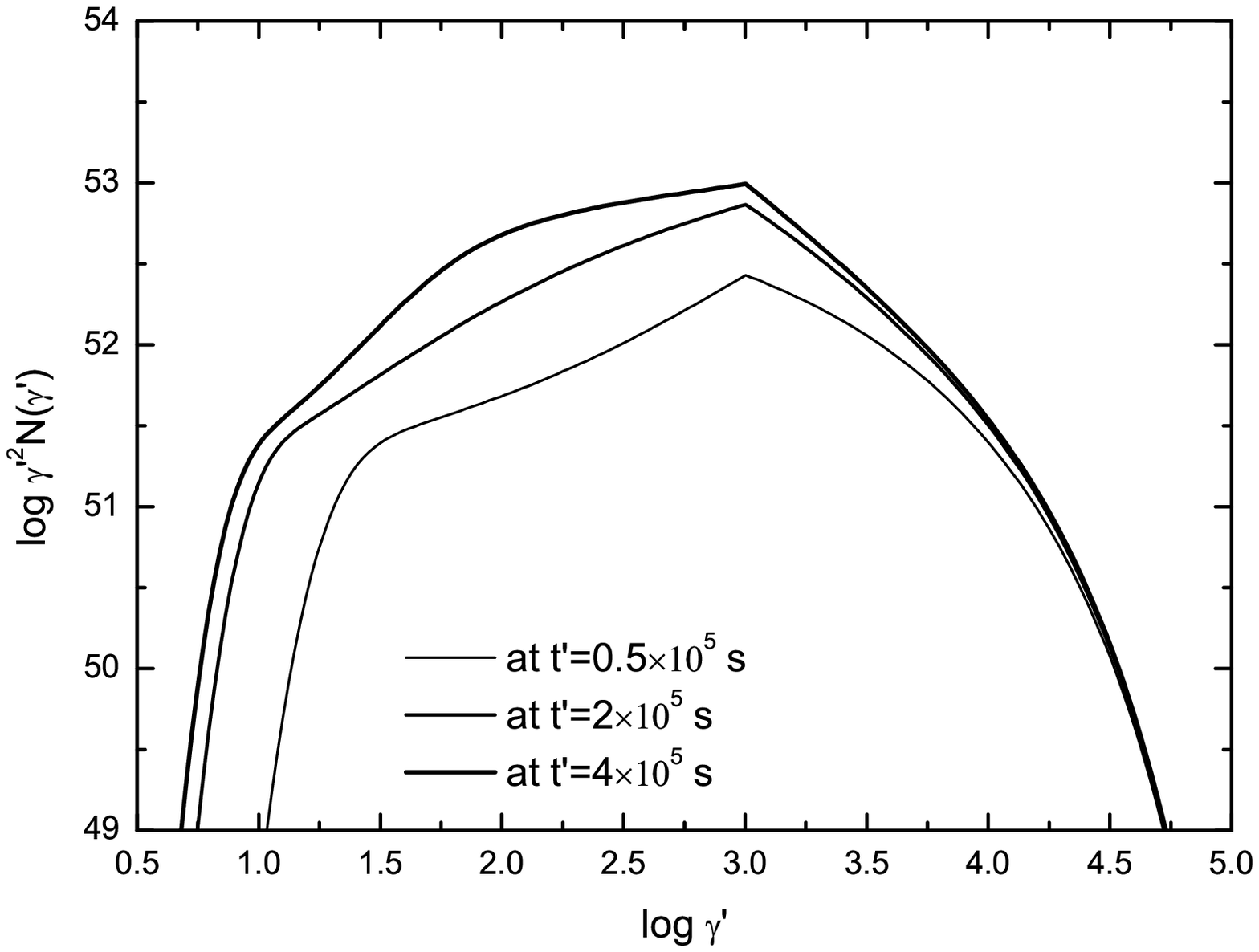}
 \includegraphics[width=270pt,height=215pt]{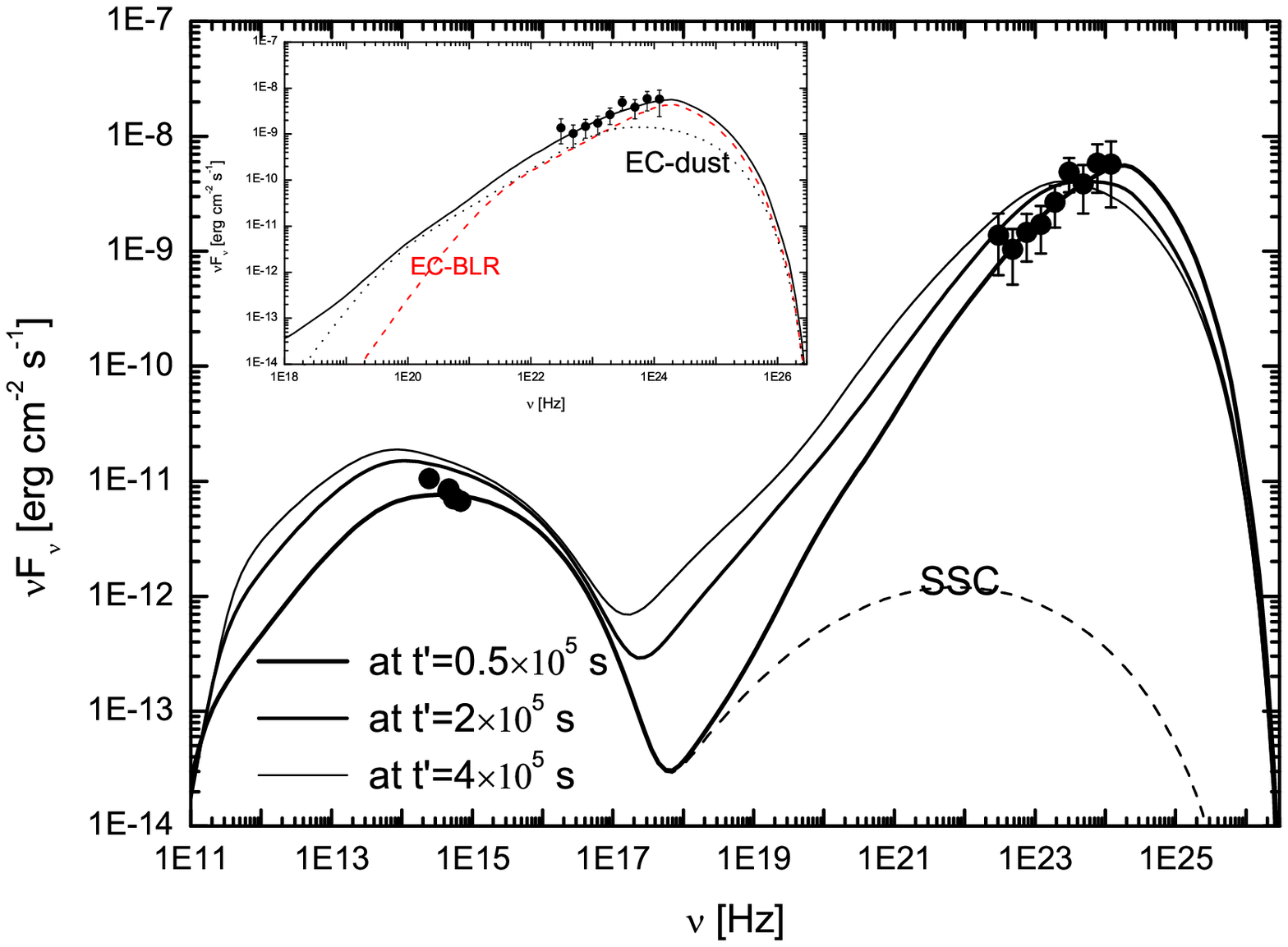}
 \caption{Modeling the very hard $\gamma$-ray spectrum of 3C 279 during 2013 December. The inset in bottom panel shows the details of modelling at gamma-ray energies. The parameters are $\tp_{\rm inj}=10^8\ $s, $\gp_{\rm min}=10^3$, $\gp_{\rm cut}=10^4$, \textbf{$B'=0.4\ $G}, $\delta_{\rm D}=42$, $R'=7.5\times10^{15}\ $cm, \textbf{$Q_0=2.5\times10^{49}\ \rm s^{-1}$, $s=2.2$, $\tau_{\rm BLR}=0.2$}, $\tau_{\rm dust}=0.3$, and $r_0=0.8\ r_{\rm BLR}$. }
\label{3c2791}
\end{center}
\end{figure}

\begin{figure} 
\begin{center}
 \includegraphics[width=270pt,height=215pt]{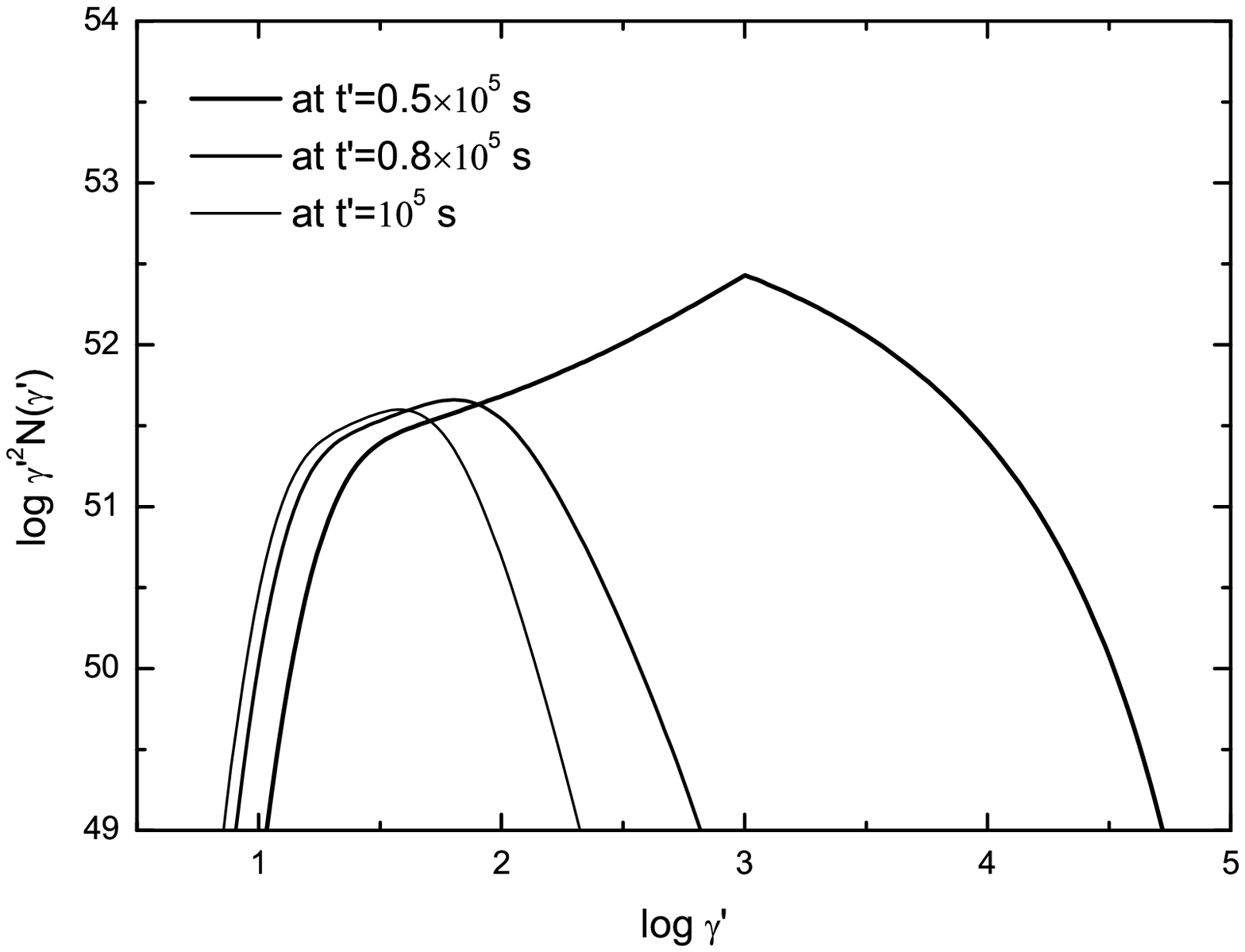}
 \includegraphics[width=270pt,height=215pt]{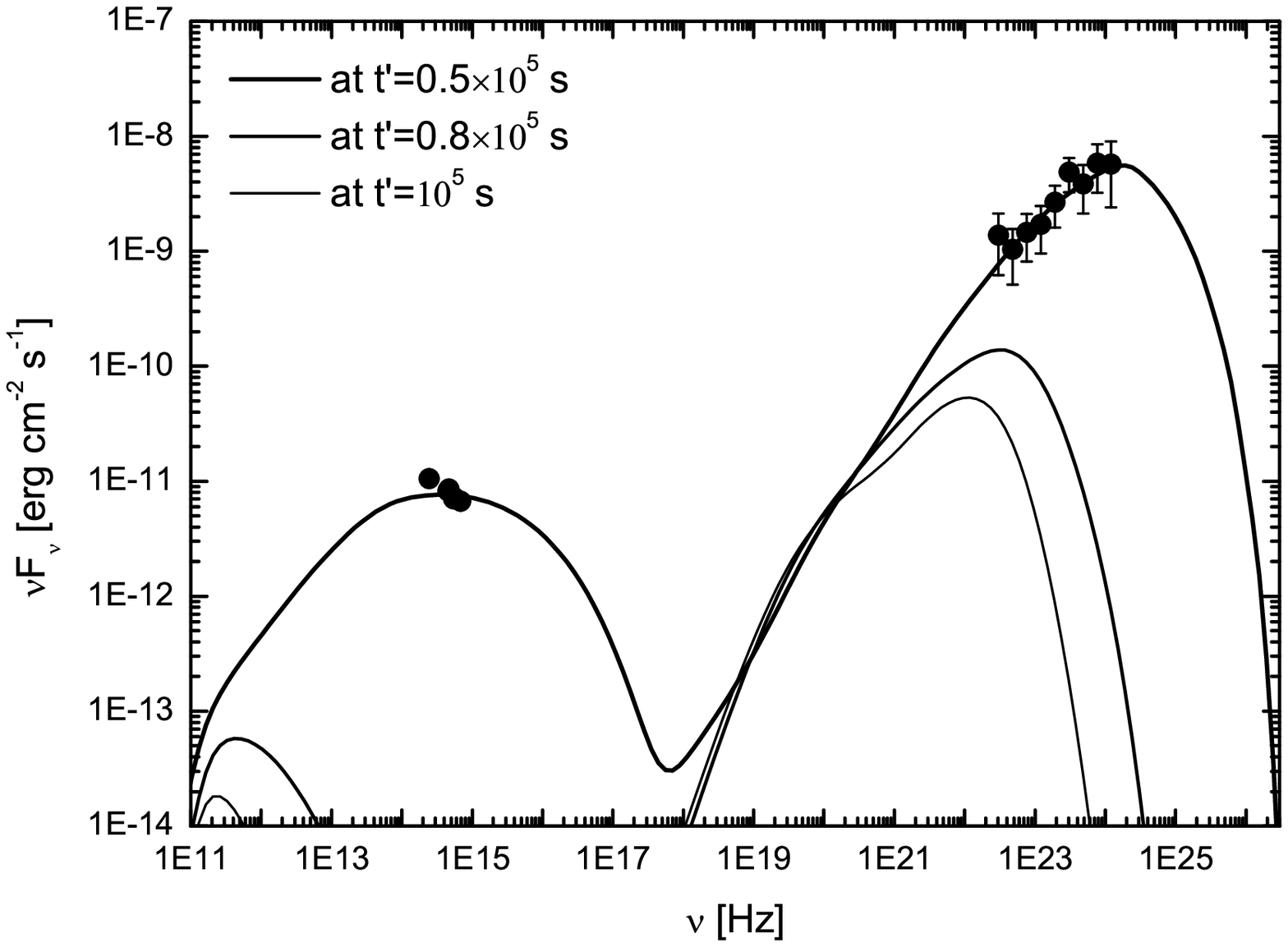}
 \caption{Same as Fig.~\ref{3c2791}, but using $\tp_{\rm inj}=0.5\times10^5\ $s. }
\label{3c2793}
\end{center}
\end{figure}

A satisfactory modelling to the very hard $\gamma$-ray spectrum as well as the optical data is seen in Fig.~\ref{3c2791};
model parameters can be found in the caption of Fig.~\ref{3c2791}.
No extreme parameter is required. The distance where the injection starts is $r_0=0.8r_{\rm BLR}$.
The magnetic field is consistent with that in other activities derived by \citet{dermer14} and \citet{yan15,yan15c} who
used a static log-parabola EED to model the SEDs of 3C 279.
The comoving frame electron injection power is $L'_{\rm inj}\thickapprox7\times10^{42}\rm \ erg \ s^{-1}$.
The ratio of emitting electron energy density (at $t'=0.5\times10^5\ $s) to magnetic energy density is $\sim$4.
The EED develops a clear form of $N'_{\rm e}(\gp)\propto\gp^{-1.3}$ below $\gp_{\rm min}$ when $\tp<10^5\ $s; as the blob moves outside the BLR ($\tp>10^5\ $s) the cooling due to EC-dust becomes relevant, and consequently a softer EED occurs.

In Fig.~\ref{3c2793}, we show the temporal evolutions of the SED and EED when $\tp>\tp_{\rm inj}=0.5\times10^5\ $s.
It is noted that the optical emission soon becomes undetectable after stopping the injection (see Fig.~\ref{3c2793}).
We note that the EED is quickly narrowed after stopping the injection.

\section{Summary and discussion}
\label{sd}

We have fully investigated the evolution of the EED in the jet of FSRQ using a time-dependent model.
We found that a very hard electron spectrum with $p\sim1.3$ below minimum injection energy is formed
in the fast-cooling scenario owing to Compton scattering BLR radiation in the KN regime.
This produces a very hard spectrum up to $\sim5\ $GeV via inverse Compton scattering BLR and dust radiations.
As we showed, our model can reasonably explain the very hard $\gamma$-ray spectrum of 3C 279 observed in the extreme flare during 2013 December. The satisfactory modelling is sensitive to the $\gamma$-ray emission site, and requires the $\gamma$-ray emission taking place inside the BLR. External $\gamma\gamma$ absorption effects are unimportant in the {\it Fermi}-LAT spectrum of 3C 279, measured below $\thickapprox$10 GeV. Absorption on BLR photons can be important above $\sim$25 GeV \citep[e.g.,][]{dermer14}.

Our model expects that X-rays lag optical and $\gamma$-ray emission, which can be tested by future observations. A more complicated injection rate might be needed to reproduce the $\gamma$-ray light-curve profile.
During the extreme $\gamma$-ray flare, the optical emission showed weak variability \citep{Hayashida15}.
The problem of lack of simultaneous optical variability might be resolved in the fast-cooling regime
where the electrons making optical emission by synchrotron radiation do not make a substantial contribution to the LAT spectrum.

The impacts of IC scattering in the KN regime on EED have been investigated by previous studies in the slow-cooling scenario \citep[e.g.,][]{dermer02,kt05,Geo06,Sikora09}.
For comparison, we also revisited this scenario.
We showed that, in the slow-cooling scenario, the electron distribution becomes harder at $\sim\gp_{\rm KN}$ with the spectral index from $\sim s+1$ ($s=2.1$ is the index of injection spectrum ) to $\sim2.5$. This moderate hardening in EED results in a flat EC-BLR/dust component, which is similar to the finding in \citet{Geo06}, also see the results in \citet{dermer02}.
We do not see a dip in $\gp^2N'_{\rm e}(\gp)$ distribution presented by \citet{kt05}. We note that \citet{kt05} also obtained a similar flat EC component. However, the formation mechanisms for such flat spectrum in the \emph{Fermi}-LAT energy range are not unique.

The difference between the cooling behaviours in the Thomson and KN regimes not only affects the EED/$\gamma$-ray spectrum, but also affects the decay of $\gamma$-ray light curve. In the KN regime, cooling time is energy-independent, while in the Thomson regime cooling time is energy-dependent. This difference has been proposed to constrain the $\gamma$-ray emission site in FSRQs \citep{Dotson12,Dotson15}.

It is interesting to note that the mean $\Gamma_{\gamma}$ for the \emph{Fermi}-LAT detected FSRQs is $2.44\pm0.20$ \citep{AGN3d}.
Analyses of the $\gamma$-ray spectra with $\Gamma_{\gamma}>2$ and the simultaneous optical and X-ray data indicated that $\gamma$-ray emissions of 3C 279 take place outside the BLR \citep{dermer14,Paliya,yan15,yan15c}.

As a last remark, we note that very recently \citet{UZ} and \citet{zhao} showed that a very hard electron spectrum with $p\sim1$ can be produced in the fast-cooling regime due to synchrotron radiation in a strongly decaying magnetic field.
This can explain the $\gamma$-ray burst (GRB) prompt emission spectra whose low-energy photon spectral index has a value $\sim1$.

\section*{Acknowledgments}
We thank the referee for a very helpful report. We are grateful to Krzysztof Nalewajko for providing us the data of 3C 279.
We thank Xiang-Yu Wang and Zhuo Li for bringing the KN effect in GRB study \citep[e.g.,][]{wang} to our attention
when this work is presented at the 8th black hole conference, held in Kunming 14 - 16th October 2015.
This work is partially supported by the National Natural Science Foundation of China (NSFC 11433004) and Top Talents Program of Yunnan Province, China. DHY acknowledges funding support by China Postdoctoral Science Foundation under grant No. 2015M570152, and by the National Natural Science Foundation of China (NSFC) under grant No. 11573026, and by Key Laboratory of Astroparticle Physics of Yunnan Province (No. 2015DG035).
 SNZ acknowledges partial funding support by 973 Program of China under grant 2014CB845802, by the National Natural Science Foundation of China (NSFC) under grant Nos. 11133002 and 11373036, by the Qianren start-up grant 292012312D1117210, and by the Strategic Priority Research Program ``The Emergence of Cosmological Structures'' of the Chinese Academy of Sciences (CAS) under grant No. XDB09000000.



\bsp	
\label{lastpage}
\end{document}